\documentclass[a4paper]{PoS}

\title{
\vspace*{-1.4cm}
\begin{minipage}{\textwidth}
{\normalfont\small TTP16-028 
\hspace{\fill} June 2016}\\
\end{minipage}\\[60pt]
Precision Measurements in Electron-Positron Annihilation: Theory  and Experiment}

\ShortTitle{Precision Measurements in Electron-Positron Annihilation}

\author{Konstantin Chetyrkin and \speaker{Johann H. K\"uhn}\\
        Institut f\"ur Theoretische Teilchenphysik, KIT, 76131 Karlsruhe\\
        E-mail: \email{Johann.Kuehn@KIT.edu}}

\abstract{
  Theory results on precision measurements in electron-positron annihilation
  at low and high energies are collected. These cover pure QCD calculations as
  well as mixed electroweak and QCD results, involving light and heavy
  quarks. The impact of QCD corrections on the $W$-boson mass is discussed
  and, last not least, the status and the perspectives for the Higgs boson
  decay rate into $b\bar b$, $c\bar c$ and into two gluons.
}

\FullConference{Loops and Legs in Quantum Field Theory\\
		24-29 April 2016\\
		Leipzig, Germany}

\begin{document}

\section{Introduction}
The determination of the strong coupling $\alpha_s$ in clean
experimental conditions is one of the important issues in ongoing
theoretical and experimental investigations. During the past years
significant progress has been made in perturbative calculations of a
large variety of processes. In this talk a number of benchmark processes
is identified and the corresponding predictions are presented to the
highest presently available order.

During the past forty years calculations in the framework of
perturbative QCD have developed from a quantitative description of a
few benchmark processes to precise predictions of numerous hadronic
processes, albeit typically at relatively high energies and/or for
inclusive reactions. Many of these are closely related to
electron-positron annihilation into hadrons, at lower energies through
the electromagnetic, at higher energies through the neutral current.
QCD corrections to the decay of the $W$-boson into hadrons through the
vector and the axial vector current can be evaluated in a similar way
and are, in turn, closely related to QCD corrections of the 
$\tau$-lepton decay rate.
The decay of the Higgs boson into hadrons, on the other hand, proceeds
through the scalar current and can be treated with very similar
methods. Finally the running of the strong coupling constant from low
energies, say $m_\tau$, up to the mass of the Higgs boson and beyond, is
governed by the beta-function, can be calculated with similar
techniques, is now available in five-loop order and will also be
discussed in this context.
\section{Electron-positron annihilation at low energies}
The cross section for electron-positron annihilation into hadrons is
well described by perturbative QCD, at least in the regions away from
the various quark thresholds. The result of the BESSII
collaboration \cite{Ablikim:2006aj}, 
 consisting of an average of measurements at   3.650~GeV and 3.6648~GeV,
\begin{equation}
\bar R=2.224\pm 0.019\pm 0.089
\end{equation}
is in good agreement with the theoretical expectation
\begin{equation}
\bar R=  3 (Q_u^2 + Q_d^2 + Q_s^2)(1+a_s+1.64010 a_s^2  - 10.28395 a_s^3
- 104.78910 a_s^3) 
\end{equation}
adopting as value of the strong coupling $\alpha_s=
0.31\pm0.14$. Although the precision of this experiment cannot
compete with those at LEP (to be discussed below), the agreement
between theory and experiment is, nevertheless, remarkable already now.
Any further improvement of the experimental precision would be welcome
and would allow the comparison of results for $\alpha_s$ at low and high energies.
Let us mention in passing, that there is in principle the (very small) singlet
contribution contribution proportional $(\sum_i Q_i)^2$, which starts 
contributing in order $\alpha_s^3$ and is also available up to order   $\alpha_s^4$.
For the three-flavour case $(\sum_i Q_i)^2$ happens to vanish, for the
four- and five-flavour case the term is numerically small
\cite{Baikov:2008jh,Baikov:2012er}.
\section{$Z$-production and -decay in electron-positron annihilation}
From the theory side there is only one slight complication when moving
from low to high energies: the axial current starts contributing and,
correspondingly, QCD corrections specific for this case start
contributing in order $\alpha_s^2$. Of course, also a singlet piece,
starting in order $\alpha_s^3$, is present, just as for the electromagnetic 
current. The corrections for the three different pieces, each evaluated
to order $\alpha_s^4$, are shown separately in Figs. 1--4.
Note that  $\alpha_s (M_Z) =0.1190$ and $n_l =5 $ are  adopted in Figs. 2--4.
\begin{figure}[!h] 
\begin{center}
\raisebox{-3.5mm}{\includegraphics[width=.27\linewidth]{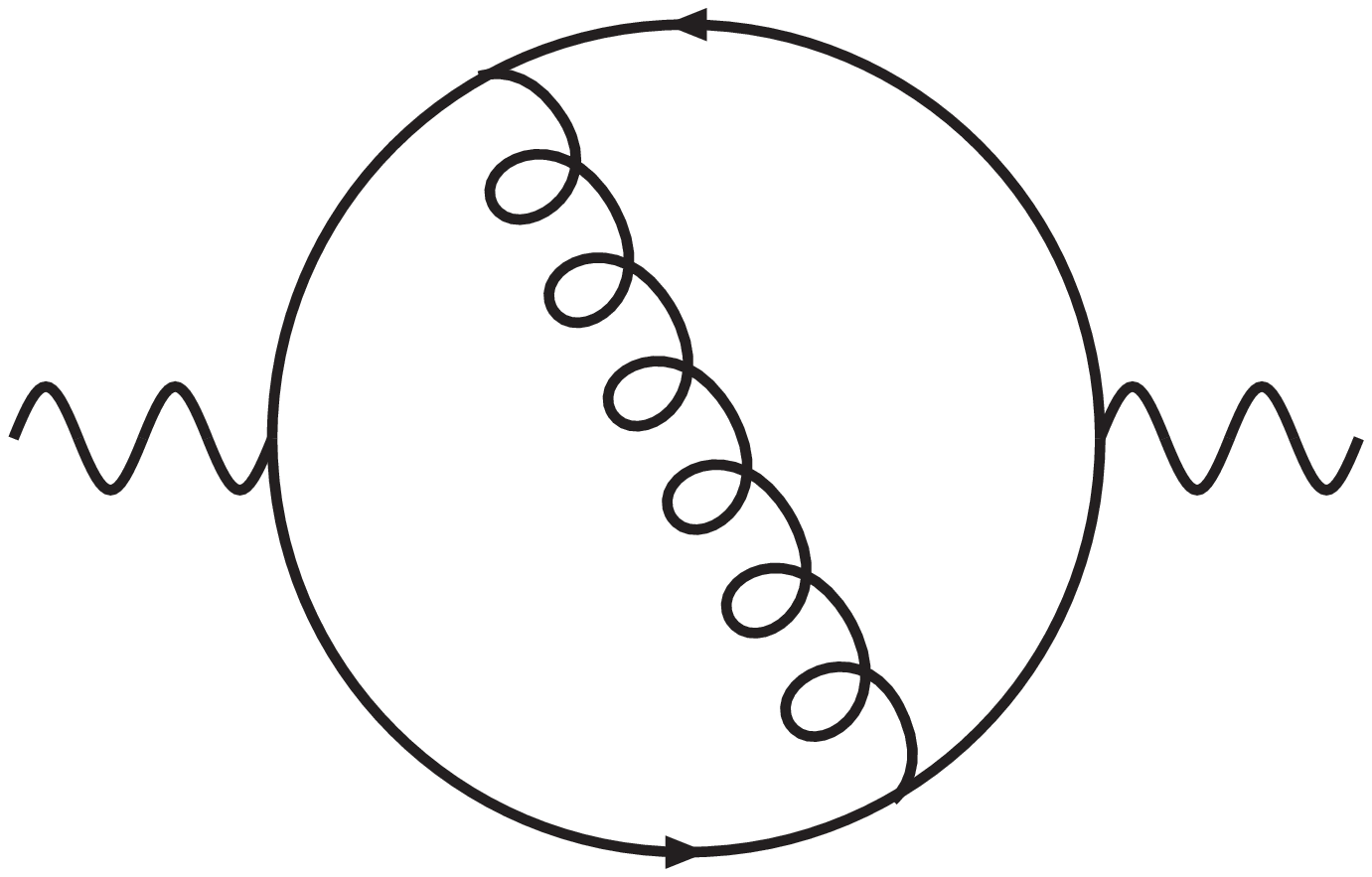}}
\includegraphics[width=.28\linewidth]{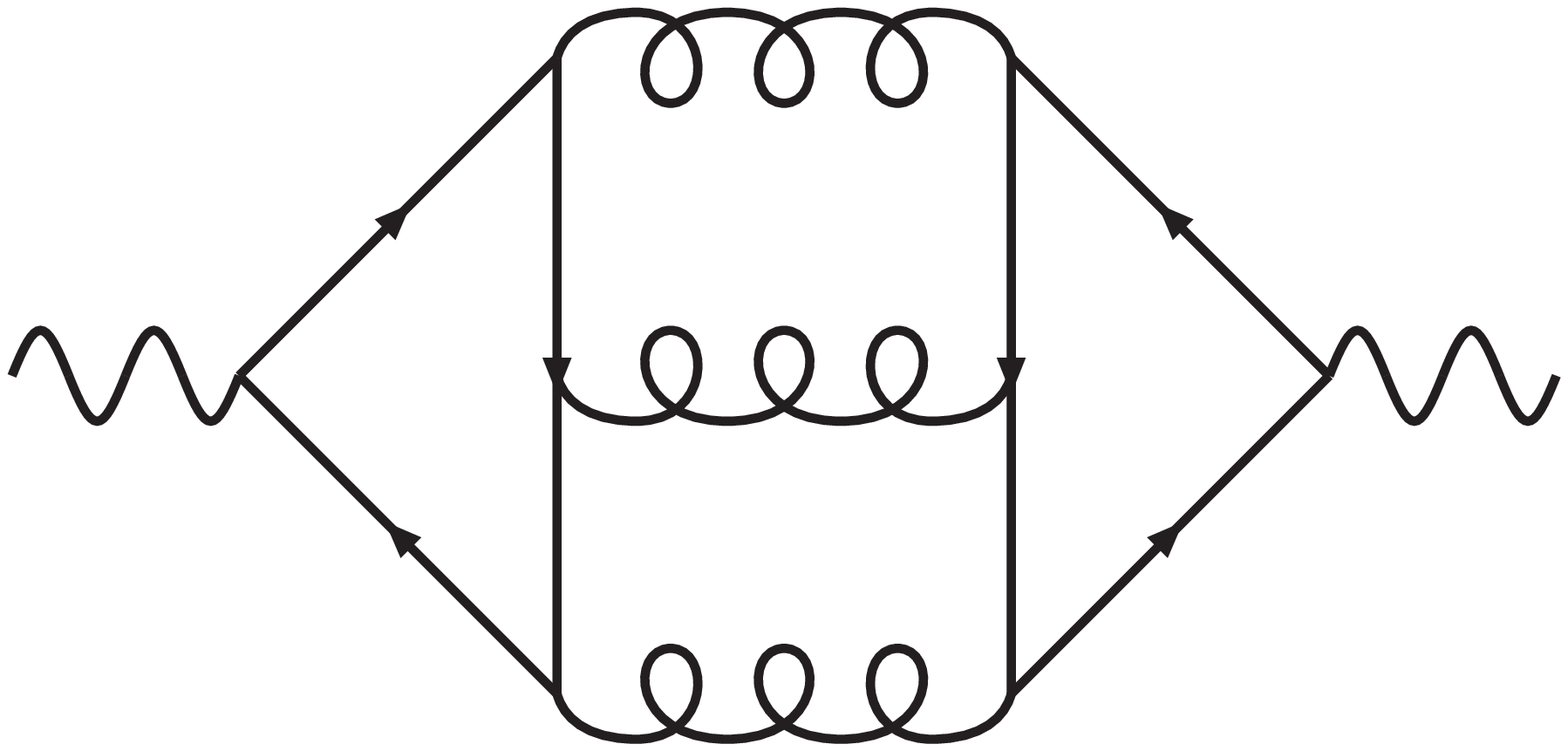}
\raisebox{-1mm}{\includegraphics[width=.28\linewidth]{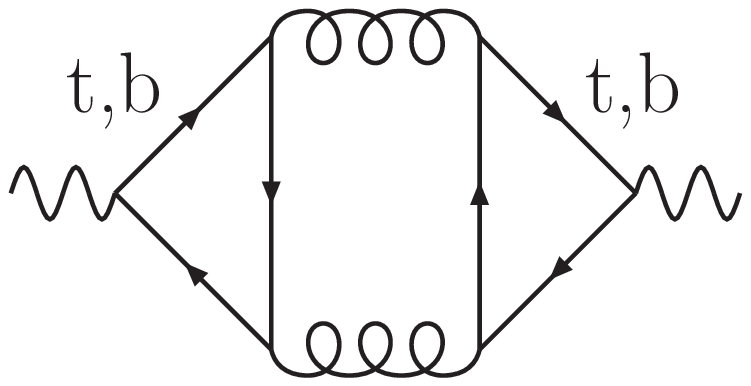}}
\end{center}
\hspace{-3.1cm} {\Large \hspace{6cm} (a) \hspace{3.1cm}  (b)  \hspace{4.cm} (c)}
\caption{{Different contributions to $r$-ratios: (a) non-singlet, (b) vector singlet and (c) axial vector singlet.}
} 
\end{figure}
\begin{figure}[!h] 
\begin{center}
\includegraphics[width=.7\linewidth]{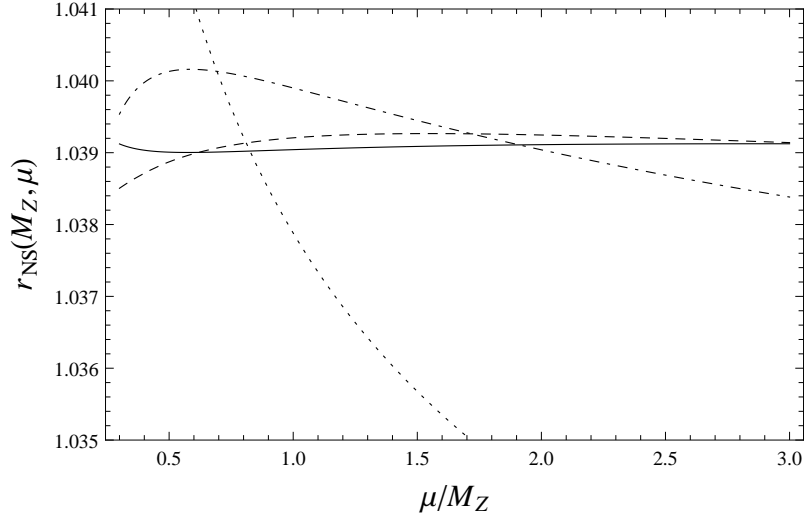}\\
\end{center}
\caption{Scale dependence of  non-singlet $r_{NS}$.
Dotted, dash-dotted, dashed and solid curves refer to $\mathcal{O}(\alpha_s)$ up to $\mathcal{O}(\alpha_s^4)$ predictions.
} 
\end{figure}
\begin{figure}[!h] 
\begin{center}
\includegraphics[width=.7\linewidth]{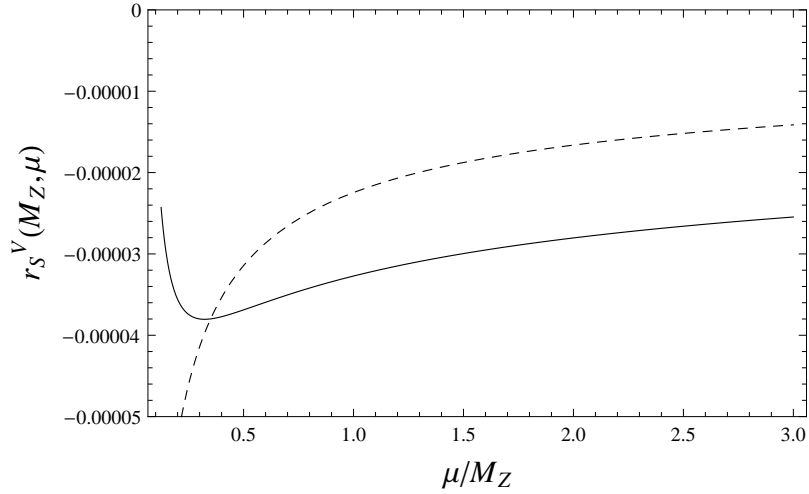}\\
\end{center}
\caption{Scale dependence of  the vector singlet $r_S^V$.
Dashed and solid curves efer to $\mathcal{O}(\alpha_s^3)$ up to $\mathcal{O}(\alpha_s^4)$ predictions.
} 
\end{figure}
\begin{figure}[!h] 
\begin{center}
\includegraphics[width=.7\linewidth]{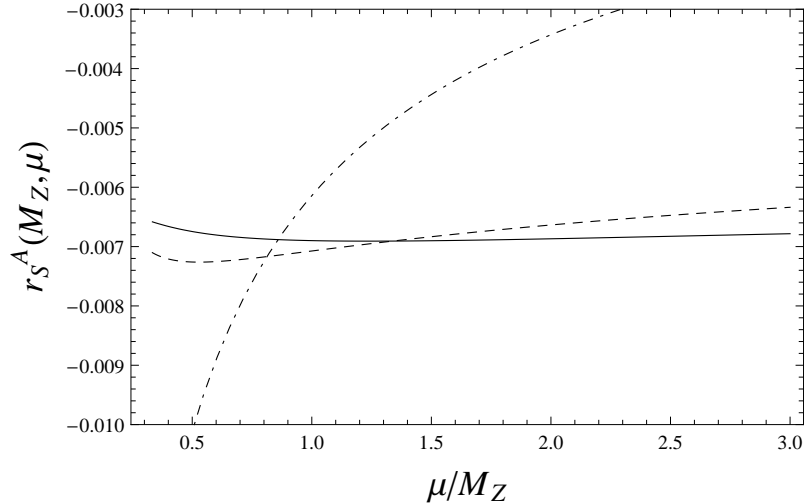}\\
\end{center}
\caption{Scale dependence of the axial vector singlet $r^A_{S;t,b}$.
Dotted, dash-dotted, dashed and solid curves refer to $\mathcal{O}(\alpha_s)$ up to $\mathcal{O}(\alpha_s^4)$ predictions.
} 
\end{figure}

\noindent
The result 
\begin{equation}
\alpha_s(M_Z)=0.1190\pm 0.0026
\end{equation}
still exhibits a sizeable error, significantly larger than the theory
error which has been estimated to \cite{Baikov:2008jh,Baikov:2012er}
$\delta \Gamma_{NS}=101 {\rm \ keV}$,
$\delta \Gamma_{S}^V=2.7 {\rm \ keV}$,
and $\delta \Gamma_{S}^A=42 {\rm \ keV}$.
Summing these errors linearly, one arrives at a theory uncertainty of
146~\ keV, which corresponds to a shift in $\alpha_s$ of about
$3\times 10^{-4}$ and is
thus about a factor ten smaller than the current experimental error,
based on $Z$ decays, $\alpha_s=0.1190 \pm 0.0026$.
\section{Mixed electroweak and QCD corrections for $Z$ decays: light and
  heavy quarks}
As a consequence of the virtual top quark one expects a significant
difference between the electroweak corrections for $Z$ decays into
$d\bar d$ and $u\bar u$ on the one hand and into $b\bar b$ on the other
hand. This pattern repeats itself  in the mixed electroweak and QCD
corrections of order $\alpha_{weak}\alpha_s$. For light quarks the
two-loop corrections of order $\alpha\alpha_s$ have been evaluated
about twenty years ago. The final result which makes the
non-factorizing terms explicit can be cast into the form \cite{Czarnecki:1996ei}
\begin{equation}
\Delta\Gamma\equiv \Gamma({\rm two~loop: EW\times QCD}) -
\Gamma_{Born}\delta_{EW}^{NLS}\delta_{QCD}^{NLO} = -0.59(3) {\rm \ MeV} 
\end{equation}
which is sufficient for the present experimental precision of 2  MeV for
the hadronic decay rate. On the other hand, given an expected
experimental precision of $\delta\Gamma \approx 0.1 {\rm \ MeV}$,
as advertised for a future electron-positron 
collider \cite{Blondel:2011fua,Gomez-Ceballos:2013zzn}, 
the next, not yet available three-loop term might eventually be required.

The situation is qualitatively similar for the $Z\to b \bar b$ decay
mode which, however, receives also contributions from virtual top
quarks.
The precision of the measured branching ratio of $15.12\pm0.05\%$ is, at
 present, quite close to the size of the two-loop term, which is given by
 \cite{Harlander:1997zb}
\begin{equation}
\Gamma_b - \Gamma_d = (-5.69 - 0.79 + 0.50 + 0.06){\rm \ MeV}  
\end{equation}
and has been split into one- and two-loop contributions and into the
$m_t^2$-enhanced piece and the rest. Let us mention in passing that part
of the three-loop corrections, the non-singlet piece, has been
evaluated in \cite{Chetyrkin:1999za}.
It amounts to about 0.1\ MeV, is irrelevant in the moment, but of potential
importance at a future electron-positron collider.

Many top-induced corrections become
significantly smaller, if the top quark mass is expressed in the
$\overline{\mbox{MS}}$ convention. The relation between pole and $\overline{\mbox{MS}}$
mass has been evaluated in  three- \cite{Chetyrkin:1999qi} and recently
even four-loop \cite{Marquard:2015qpa} approximation and reads
\begin{equation}
\bar m_t(\bar m_t) = m_{\rm pole} 
(1-1.33  \, a_s -  6.46 \, a_s^2 - 60.27 \, a_s^3 -704.28 \, a_s^4)
=(163.45 \pm 0.72|_{m_t} \pm 0.19 |_{\alpha_s} \pm ?|_{th}) {\rm \ GeV}
\end{equation}
with a theory error of about 100~MeV.

\section{The $W$ boson mass from $G_F$, $M_Z$, $\alpha$ and the rest}
The present precision \cite{Agashe:2014kda}
 of $M_W=80.385\pm 0.015\ {\rm \ MeV}$ is based on a combination
of LEP, TEVATRON and LHC results. In contrast, at a future linear or
circular electron-positron collider a precision better than 1 MeV is
advertised \cite{Blondel:2011fua,Gomez-Ceballos:2013zzn}.
In Born approximation the $W$ boson mass can
be derived from the Fermi coupling $G_F$, the $Z$ boson mass and the
electromagnetic coupling $\alpha$. The rest of the parameters, in
particular the masses of fermions and the Higgs boson, enter
through radiative corrections. Numerically one finds for the shift
in the $W$-boson mass induced by virtual contributions of the top quark
\begin{equation}
\delta M_W\approx \frac{1}{2} M_W
\frac{\cos^2\theta_W}{\cos^2\theta_W - \sin^2\theta_W} \approx 5.7 \times
    10^4 \delta \rho [(\rm MeV)]
{},
\end{equation}
with the $\rho$ parameter calculated in three-\cite{Chetyrkin:1995ix,Avdeev:1994db} and even
four-loop \cite{Chetyrkin:2006bj,Boughezal:2006xk}
approximation
\begin{equation}
\delta \rho_t = 3 X_t (1-2.8599 \, a_s -14.594 \, a_s^2 - 93.1 \, a_s^3)
\end{equation}
The three- and four-loop terms correspond  to shifts of 
$\delta M_W=9.5 {\rm \ MeV}$ and $\delta M_W=2.1 {\rm \ MeV}$
respectively. The three-loop term is quite comparable to the current
experimental sensitivity, the four-loop term would become relevant at a
future electron-positron collider.


At this point it should be emphasized that in three-loop approximation a
variety of mixed QCD and electroweak corrections are
available \cite{Faisst:2003px}, which amount to 2.5~MeV for the mixed
terms proportional $\alpha_s X_t^2$ and to 0.2~MeV for the purely weak
terms of order $X_t^3$. While these are certainly below the anticipated
experimental precision for the near future, they might well become
relevant at a future $e^+e^-$ collider. At the same time a number of not
yet calculated terms might eventually become relevant, for example
four-loop tadpoles of order $\alpha_s^2X_t^2$ or even five-loop terms of
order $\alpha_s^4X_t$. Although not yet relevant for the moment,
these corrections might well enter the analysis of experiments at a future
linear or circular $e^+e^-$ collider. 

Let us also mention that many corrections are significantly smaller if
the top quark mass is expressed in terms of the $\overline{\mbox{MS}}$-mass, or
closely related quantities, 
like the potential subtracted (PS)~\cite{Beneke:1998rk},
1S~\cite{Hoang:1998hm,Hoang:1998ng,Hoang:1999zc} or renormalon subtracted
(RS)~\cite{Pineda:2001zq} one.
In other words, a large part of
the corrections can be absorbed  in the relation between the 
$\overline{\mbox{MS}}$- and the pole mass, discussed above.
Let us emphasize that e.g. the potential subtracted top quark mass (and
as well as other ``short-distance'' masess) could be determined at
electron-positron colliders with a significantly higher precision,
reaching 20 to 30 MeV.

The present, relatively large experimental error in the top mass is
necessarily connected to its determination at a hadron collider. The
situation would be significantly better at an $e^+e^-$ machine, where
uncertainties around or even below 50~MeV might be possible\cite{Baer:2013cma},
and even 10 to 20~MeV have been quoted 
\cite{Blondel:2011fua,Gomez-Ceballos:2013zzn}.

Let us mention in passing that the total cross section for electron-positron
annihilation into hadrons at low energies, below the $Z$ resonance, 
receives QED corrections connecting initial
and final state in order $\alpha^2$ and hence two loop only. This is a
consequence of Yang's theorem  which forbids contributions from
triangular fermion graphs. This is different in the full electroweak
theory, where mixed triangular contributions with vector and axial
vector couplings start to contribute in one-loop approximation already. 
In addition there is a huge tail from  ISR QED corrections
which increases the cross section by about a factor three and must be
carefully controlled to  achieve a realistic result for the $R$ ratio.

\section{Perspectives for $e^+e^-\to Z + H(\to hadrons)$}
One of the most important reactions at a future electron-positron
collider will be the production of the Higgs boson in the process
$e^+e^-\to Z + H$ with the subsequent decay of the Higgs boson into
hadrons, i.e. quarks and gluons.
Let us demonstrate the status of recent calculations in a few selected examples:

The Higgs boson decay into bottom-antibottom quarks is of course governed
by the mass of the bottom quark, evaluated at the scale of $m_H$. In
total the rate is given by \cite{Baikov:2005rw}
\begin{equation}
\Gamma(H\to b \bar{b})=
\frac{G_F M_H}{4\sqrt{2} \pi} m_b^2(\mu^2(M_H^2)) R_S(s=M_H^2, \mu^2)
\end{equation}
with
\begin{eqnarray}
R_S(s = M_H^2, \mu^2=M_H^2)
& = & 1+ 5.667 \frac{\alpha_s}{\pi} + 29.147  \frac{\alpha_s}{\pi}^2 +
41.758 \frac{\alpha_s}{\pi}^3 - 825.7  \frac{\alpha_s}{\pi}^4 \\
 & = & 1 + 0.1948 + 0.03444 + 0.0017 - 0.0012 = 1.2298  
\end{eqnarray}
Here $\alpha_s=\alpha_s(M_H)=0.108$, corresponding to
$\alpha_s(M_z)=0.118$ has been adopted.
The decay rate depends on two phenomenological parameters, the strong
coupling and the bottom quark mass. To avoid the appearance of large
logarithms of the type $\ln (\mu^2/M_H^2)$, the parameter $\mu$ should
be chosen around $M_H$. However, the starting value of $m_b$ is
typically determined at much smaller values, typically around 5 to
10~GeV \cite{Chetyrkin:2009fv}.
 The evolution from this low scale to $\mu=M_H$
is governed by the quark mass anomalous dimension$\gamma_m$ and the
$\beta$ function, both of which must be known in five-loop order
\cite{Baikov:2014pja,Baikov:2016tgj}
in order to match the accuracy of the fixed order result. 
For the quark mass value 
$m_b(10 {\rm GeV}) = 3610 -(\frac{\alpha_s(M_Z) - 0.118}{0.002})^2 \times 12
\pm 11 \ {\rm MeV}$ 
one finds 
$m_b(M_H)= 2759 \pm 8|_{m_b} \pm 27|_{\alpha_s}\,\,{\rm MeV}$.
The remaining theory uncertainty from our ignorance of higher order
corrections amounts to about 1.5 permille and is completely negligible.

Let us list the potential improvements which might develop during the
coming years: The strong coupling constant might be known to
$\delta\alpha_s(M_Z)=2\times 10^{-4}$ and the bottom quark mass with a
relative precision of $\delta m_b / m_b \approx 10^{-3}$. In total this
would lead to a relative precision 
\begin{equation}
\frac{\delta \Gamma(H\to b\bar b)}{\Gamma(H\to b\bar b)}=
\pm 2   \times 10^{-3}|_{m_b} 
\pm 1.3\times 10^{-3}|_{\alpha_s}
\pm 1   \times 10^{-3}|_{theory}
\end{equation}
which corresponds to a dramatic improvement compared to present theory
estimates.

Similar statements do apply for the $H\to c\bar c$ mode with its rate
being smaller by about a factor $((m_c(M_H) / m_b(M_H))^2 $. In this
case the reduction of $\delta m_c(3 {\rm GeV})$ from 13~MeV to 5~MeV seems
conceivable, reducing the uncertainty from 
$\delta m_c(3{\rm GeV}) / m_c(3{\rm GeV}) = 13 {\rm \ MeV} / 986  {\rm \ MeV}$
to $5 {\rm \ MeV} / 986  {\rm \ MeV}$.
At the scale of $M_H$ this would lead to a reduction of the error in
$m_c(M_H)$ from $m_c(M_H) =( 609 \pm 8|_{m_c} \pm 9|_{\alpha_s}) {\rm  \ MeV}$
to $\pm 3 {\rm \ MeV}$. This, in turn, would lead to a reduction of the
relative error of $\delta \Gamma (H\to c\bar c)/\Gamma (H\to c\bar c)$
from $5.5\times 10^{-2}$ to $1\times 10^{-2}$. In absolute terms the
errors of $H\to c\bar c$ and $H\to b\bar b$ are then compatible.

Finally, let us briefly mention another prominent decay mode of the
Higgs boson, its decay into two gluons, which is available in order
$\alpha_s^5$ and given by \cite{Baikov:2006ch} 
\begin{equation}
\Gamma(H\to gg) = K \Gamma_{Born}(H\to gg)
\end{equation}
with
\begin{equation}
K=1+17.9167 a_s + (156.81 - 5.71 \ln \frac{M_t^2}{M_H^2})\, a_s^2 
+(467.68 - 122.44   \ln \frac{M_t^2}{M_H^2} +10.94  \ln^2 \frac{M_t^2}{M_H^2})
\, a_s^3
{}.
\end{equation}
For the specific choice $M_t=175{\rm \ GeV}$, $M_H=125 {\rm \ GeV}$ and 
$a_s=\alpha_s^{(5)}(M_t)/\pi = 0.0363$ one finds a correction factor
\begin{equation}
K= 1+17.9167 \, a_s + 152.5 \, a_s^2 +381.5 \, a_s^3=
1+0.65038+0.20095+0.01825=1.86957
\end{equation}
Considering the claim that the experimental precision at a future electron-positron
collider might reach 1.4\%, experimental and theoretical uncertainties
would match nicely.

Although the decay of the Higgs boson into photons constitutes only a
small fraction of events, this is partly compensated by the fact
that these events are particularly clean and thus can be dug out from a
huge background. The one- and two-loop corrections 
can be written in the form \cite{Maierhofer:2012vv}
\begin{eqnarray}
\Gamma(H\to\gamma\gamma) = \frac{M_H^3}{64\pi}(A_{\rm LO}^2 +
\frac{\alpha}{\pi}(2A_{\rm LO}\, A_{\rm NLO-EW}) &+&
\frac{\alpha_s}{\pi} (2 A_{\rm LO}\,  A_{\rm NLO-QCD}) 
\\
&+&
\frac{\alpha_s}{\pi}^2 (2 A_{\rm LO} \, {\rm Re}(A_{\rm NNLO}) +
A_{\rm NLO}^2))
{},
\end{eqnarray}
where the two-loop electroweak correction was taken from
 \cite{Passarino:2007fp}.
For the actual values $M_H=126 {\rm  \ GeV}$, $m_t(M_H)=166 {\rm \ GeV}$ and $\alpha_s(M_H)/\pi=0.0358$ one finds
\begin{equation}
\Gamma(H\to\gamma\gamma)=(9.398\times 10^{-6} - 1.48\times 10^{-7}
    +1.68\times 10^{-7} + 7.93\times 10^{-9}){\rm \ GeV}
=9.425\times 10^{-6} {\rm \ GeV}
{},
\end{equation}
where the four terms describe Born approximation, electroweak
correction, QCD correction and order $\alpha_s$ and order $\alpha_s^2$
respectively.
Upon closer inspection one finds that this prediction is good to about
one permille, which should be sufficient in the foreseeable future.

The work   was supported by the Deutsche
Forschungsgemeinschaft through CH1479/1-1.

\end{document}